\documentclass{iopart}
\usepackage{iopams}
\usepackage{graphicx}
\usepackage{units}
\usepackage{subfig}
\usepackage{ulem}
\usepackage{color}
\usepackage[latin1]{inputenc}
\usepackage{cite}
\usepackage{upgreek}

\begin{document}

\title{Evaluation of heat extraction through sapphire fibers for the GW observatory KAGRA}

\author{A. Khalaidovski$^1$, G. Hofmann$^2$, D. Chen$^1$, J. Komma$^2$, C. Schwarz$^2$, C. Tokoku$^1$, N. Kimura$^3$, T. Suzuki$^3$, A. O. Scheie$^4$, E. Majorana$^5$, R. Nawrodt$^2$, and K. Yamamoto$^1$}

\address{$^1$  Institute for Cosmic Ray Research (ICRR), The University of Tokyo, 5-1-5, Kashiwanoha, Kashiwa, Chiba, 277-8582, Japan\\
$^2$ Friedrich-Schiller-Universit\"at, Institut f\"ur Festk\"orperphysik, Helmholtzweg 5, D-07743 Jena, Germany\\
$^3$ High Energy Accelerator Research Organization (KEK), 1-1 Oho, Tsukuba, Ibaraki, 305-0801, Japan\\
$^4$ Grove City College, 100 Campus Dr, Grove City, PA 16127, USA\\
$^5$ INFN, Sezione di Roma 1, I-00185 Rome, Italy}
\ead{yamak@icrr.u-tokyo.ac.jp}

\date{\today}

\begin{abstract}
Currently, the Japanese gravitational wave laser interferometer KAGRA is under construction in the Kamioka mine. As one main feature, it will employ sapphire mirrors operated at a temperature of 20\,K to reduce the impact from thermal noise. To reduce seismic noise, the mirrors will also be suspended from multi-stage pendulums. Thus the heat load deposited in the mirrors by absorption of the circulating laser light as well as heat load from thermal radiation will need to be extracted through the last suspension stage. This stage will consist of four thin sapphire fibers with larger heads necessary to connect the fibers to both the mirror and the upper stage. In this paper, we discuss heat conductivity measurements on different fiber candidates. While all fibers had a diameter of 1.6\,mm, different surface treatments and approaches to attach the heads were analyzed. Our measurements show that fibers fulfilling the basic KAGRA heat conductivity requirement of $\kappa\geq\,$5000\,W/m/K at 20\,K are technologically feasible.
\end{abstract}

%

\section{Introduction}
\label{sec:intro}
At the present moment, the kilometer-scale laser-interferometric gravitational wave (GW) observatories LIGO~\cite{ligo} in the USA and Virgo~\cite{virgo} in Italy are undergoing an extensive upgrade to the so-called \textit{second} or \textit{advanced} observatory generation. The installation of upgrades into the German-British observatory GEO\,600, that has a shorter baseline length but is equipped with the unique technique of quantum noise reduction using squeezed-light~\cite{khalaidovski12,grote13}, has been completed recently. In Japan, the second generation observatory KAGRA~\cite{kagra} (initially called LCGT~\cite{kuroda99}) is currently under construction, aiming at a detection of gravitational waves within the present decade. To reach its design sensitivity, KAGRA will, for the first time in long-baseline GW observatories, employ interferometer mirrors operated at cryogenic temperatures and be constructed underground in the Kamioka mine, located 220\,km west from Tokyo. The cryogenic operation will allow a reduction of the thermal noise, being the dominant noise source in the main observation frequency band of a few tens up to a few hundreds Hz~\cite{kagra,somiya13}; meanwhile, the underground location~\cite{sato04} and the employment of an advanced anti-vibration technique~\cite{takahashi08} will make it possible to strongly attenuate seismic noise at low frequencies. Both approaches are of a vital interest not only for KAGRA but also for future generations of GW observatories, e.g. being crucial for the European 3rd-generation GW observatory Einstein Telescope (ET) for which a design study was recently presented~\cite{et}.

To sufficiently reduce the thermal noise, it is planned to operate the main interferometer mirrors, i.\,e.~the input test masses (IMT) and the end test masses (ETM), at a temperature below 20\,K. Any radiation absorbed in the test masses will have to be extracted via 4 sapphire fibers with a length of 300\,mm and a diameter of 1.6\,mm. The upper end of the fiber, fixed to the intermediate test mass, will be kept at a temperature of 16\,K. From there, the heat will be extracted through pure-metal flexible heat links connected to an upper platform and then to the inner radiation shield. For a detailed discussion of the layout of the KAGRA cryogenic system, please see~\cite{kagra} and references therein. 

In general, the amount of heat that can be transferred through a cylinder of diameter $d$, length $l$ and the temperatures $T_1$ and $T_2$ at both ends of the link can be calculated via
\begin{equation}
K = \int_{T_1}^{T_2}\frac{\pi d^2}{4l}\kappa(d,T)\textrm{d}T,
\label{equation1}
\end{equation}
where $\kappa(d,T)$ is the thermal conductivity that is a function of temperature $T$, diameter $d$, intrinsic material properties as well as of the surface roughness. For KAGRA, a minimal necessary value of $\kappa(d=1.6\,\textrm{mm},T=20\,\textrm{K})\geq5000$\,W/m/K was formulated~\cite{yamamoto13}. As discussed in~\cite{kagra}, initially up to 80\,W of laser power will be available for injection into the interferometer. Taking into account the power-recycling gain of 10 and the intra-cavity power buildup, being about a factor of 1000, this means that a maximal laser power of 800\,W would be transmitted through the ITM and thus a maximal intra-cavity power of 400\,kW. It is worth pointing out that sapphire absorption values reported at the wavelength of 1064\,nm over the last two decades exhibit a significant variance. In~\cite{blair97}, five different samples were analyzed, finding absorption values to range from 3.1\,ppm/cm to about 200\,ppm/cm at room temperature. In~\cite{tomaru01}, Tomaru \textit{et al.}~measured absorption at 1064\,nm and 5\,K, finding a mean value of 90\,ppm/cm. Latest measurements on HEM sapphire (best quality available commercially at the time of measurement) that shall also be used as KAGRA test mass material have shown a spread of 30\,ppm/cm to 70\,ppm/cm, with one sample having a much higher absorption due to unknown reasons (possibly damaged)~\cite{hemex,hirose11}. Assuming an optimistic scenario of an absorption of $\alpha_\textrm{mirr}=30$\,ppm/cm in the sapphire mirrors, and an absorption of $\alpha_\textrm{coat}=0.5$\,ppm in the dielectric SiO$_2$/Ta$_2$O$_5$ mirror coatings~\cite{yamamoto06}, 550\,mW of power will be absorbed in the ITM and have to be extracted via four 1.6\,mm thick fibers. In addition, 200\,mW are estimated to be introduced through heat radiation via the apertures for the laser beam in the cryostat. However, only a small fraction of this heat load (on the order of 10\,mW) will be absorbed by the mirror itself~\cite{sakakibara12}, so that it is negligible for the discussion in this paper. In the case of a slightly more conservative estimation of $\alpha_\textrm{mirr}=50$\,ppm/cm and $\alpha_\textrm{coat}=1$\,ppm, in total up to 1000\,mW could be deposited in the ITM, so that the laser power would need to be decreased to about 60\,\% of the maximal value to ensure the ITM temperature to remain below 20\,K. Thus, the laser power finally used will depend on a set of material parameters yet unknown that can only be determined after manufacturing and characterization of the test masses.\\

Up to now, research on the thermal conductivity of sapphire at cryogenic temperatures was mainly constrained to bulk samples. In 1955, Berman \textit{et al.} demonstrated that above temperatures of about 55\,K the heat transport in the crystal is constrained by phonon-phonon scattering and thus by intrinsic crystal properties~\cite{berman55}. In contrast, the heat flow at low temperatures is determined by phonon scattering at the crystal boundaries. Thus, in this temperature regime (that is also the one of interest for KAGRA), the thermal conductivity strongly depends on the surface treatment of the samples, and, for small diameters, also on the sample diameter. Thus, for pure crystals the effective thermal conductivity shows a steep increase, reaching a maximum between 30\,K and 50\,K, as depending on the sample, and then converges to the bulk values. Berman \textit{et al.} analyzed samples with diameters of 1\,mm to 3\,mm. For a sample with 1.6\,mm diameter and no surface treatment, a value of $\kappa(20\,\textrm{K})\approx2$\,kW/m/K was found. Higher values of 5\,kW/m/K\,-\,10\,kW/m/K were found for samples of 2.5\,mm diameter that were annealed and flame-polished. In 2002, Tomaru \textit{et al.} analyzed the heat conductivity of thin sapphire fibers of 160\,$\upmu$m, 250\,$\upmu$m and 390\,$\upmu$m diameter in view of a possible application in LCGT~\cite{tomaru02}. They reported values ranging from 550\,W/m/K to 2\,kW/m/K at a temperature of 20\,K. Also, they have shown that by damaging the fiber surface, a drop in heat conductivity by a factor of 2 could be induced. Up to now, proof that fibers with a diameter of 1.6\,mm which meet the KAGRA requirement of $\kappa_\textrm{req}(20\,\textrm{K})\geq 5$\,kW/m/K are technically feasible remained a desideratum.

Another boundary condition is the fiber geometry. The analysis in~\cite{tomaru02} has shown that a system where the mirror is suspended in two sapphire fiber loops is not feasible in terms of heat extraction, since fibers with a cross section high enough to reach sufficient thermal conductivity values would be too stiff and break in an attempt to reach a radius of 110\,mm as required for the KAGRA test masses (for a discussion of the elastic bending limit of sapphire fibers please see Fig.~2 of reference \cite{tong00}). 
\begin{figure}[b]
	\centering
\includegraphics[width=10cm]{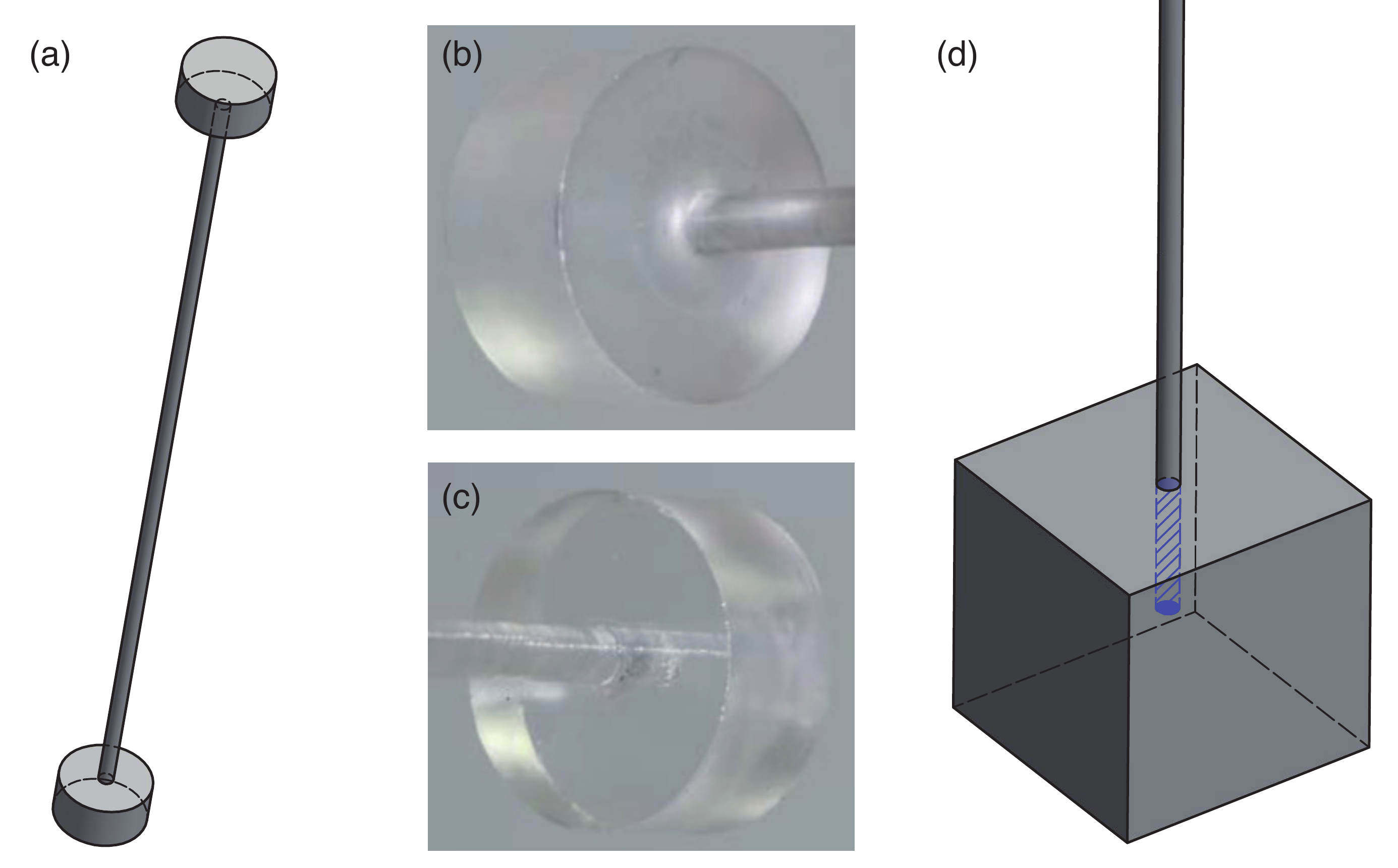}
		\caption{(a) CAD drawing of a first prototype KAGRA suspension fiber. To facilitate handling, the fibers have with 100\,mm only 1/3 of the full length, while the KAGRA design diameter of 1.6\,mm was preserved. The nail heads measure 10\,mm in diameter and 5\,mm in height. (b,\,c) Photographs of a monolithic (b) and a welded (c) fiber - nail head connection. For the latter, a plasma-welding procedure was applied. (d) Zoomed view of a more advanced nail head design, featuring up to 6 surfaces that can be used for contacting the fiber to other parts of the suspension. The welded part, having a height of 10\,mm, is highlighted by the blue shaded area.}
\label{fig:fiber}
\end{figure}
Therefore, a fiber with two cylindrical nail heads, as shown in Fig.~\ref{fig:fiber}\,(a), was proposed. Several methods of attaching this nail head to the test mass, such as hydroxide catalysis bonding (for details, see \cite{douglas14} and the references therein, \cite{dari10,suzuki06}), direct optical contact~\cite{twyford} or brazing using a thin metal layer are currently under investigation. Here, indium seems to be a very promising material. A first detailed investigation at room temperature in view of an application for GW astronomy was given in~\cite{twyford}. In~\cite{braginsky} it was shown that employing indium as coupling material in the assembly of a fused silica torsion pendulum mount allowed the measurement of a mechanical Q exceeding a value of $10^8$~\cite{mitrofanov}.  

A requirement common to all approaches is that a high flatness of the contact surfaces, being in the order of ($\lambda=632\,$nm)$/10$, is beneficial. At the moment two methods can be used to manufacture such fibers. One is to grind a fiber out of a sapphire rod (in the following called \textit{monolithic}). In the second procedure, two nail heads that are previously polished and a fiber rod of 1.6\,mm diameter are produced separately. This rod is then placed in corresponding holes drilled into the heads and the system is welded using a plasma-welding technique. These fibers are provided by the company IMPEX~\cite{impex}; photographs of a monolithic and a welded nail head are shown in Fig.~\ref{fig:fiber}\,(b,c). Only in welded fibers it was to date possible to produce nail head surfaces with a flatness better than $\lambda/10$. To keep the layout more flexible it seems convenient to adapt the nail head design to a cubic shape as shown in Fig.~\ref{fig:fiber}\,(d). Thus, more contact surfaces would be available. A crucial question in the choice of the manufacturing technique (monolithic vs.~welded) is, however, whether the weld constitutes a thermal barrier or might introduce additional mechanical loss that would lead to an increase of the suspension thermal noise.

In this paper, we present thermal conductivity measurements on prototype KAGRA sapphire fibers with a diameter of 1.6\,mm and different surface treatments. Both monolithic and welded fibers were investigated. Finally, the heat extraction through monolithic and welded fiber nail heads in view of a possible thermal resistance was compared.

\section{Heat conductivity measurements in prototype KAGRA fibers}
\label{sec:Tc}
In a first set of measurements, the thermal conductivity of different fiber types was investigated. The following samples, all provided by the company IMPEX~\cite{impex} (the only manufacturer known to us that can satisfy the design requirements), were analyzed:
\begin{enumerate}
\item For the first sample, a cylindrical nail head (diameter 10\,mm, height 5\,mm) with a neck of about 10\,mm in height and 1.6\,mm in diameter was machined out of a single sapphire crystal. Separately, the fiber rods of 80\,mm - 90\,mm length were produced using the Stepanov method~\cite{antonov} and connected to the neck. Finally, a second nail head was prepared separately with a central drilling in which the fiber rod was inserted and attached using a plasma-welding procedure, creating a quasi-monolithic connection. A visual inspection has shown that fibers produced by this approach have a notably inhomogeneous surface. These two samples were analyzed. They are referred to as fibers (a) and (b) in this paper.
\item To aid comparison, fibers of an identical geometry were ground out of a single sapphire crystal rod. No additional surface treatment was applied. These fibers are addressed as samples (1\,-\,4) in the following.
\item Finally, a rod with diameter of 1.6\,mm was produced from HEM sapphire. After grinding it was annealed at a temperature of 1700\,°C under protective atmosphere (a procedure referred to as \textit{thermo-polishing} by the manufacturer) for one hour to reduce surface defects, after which it was allowed to slowly cool to room temperature. Two nail heads were produced separately. Finally, the fiber was joined with the heads using plasma-welding, as in samples (a) and (b), to form a quasi-monolithic structure. It is addressed as fiber (i) in the paper. 
\end{enumerate}

\begin{figure}
	\centering
\includegraphics[width=12cm]{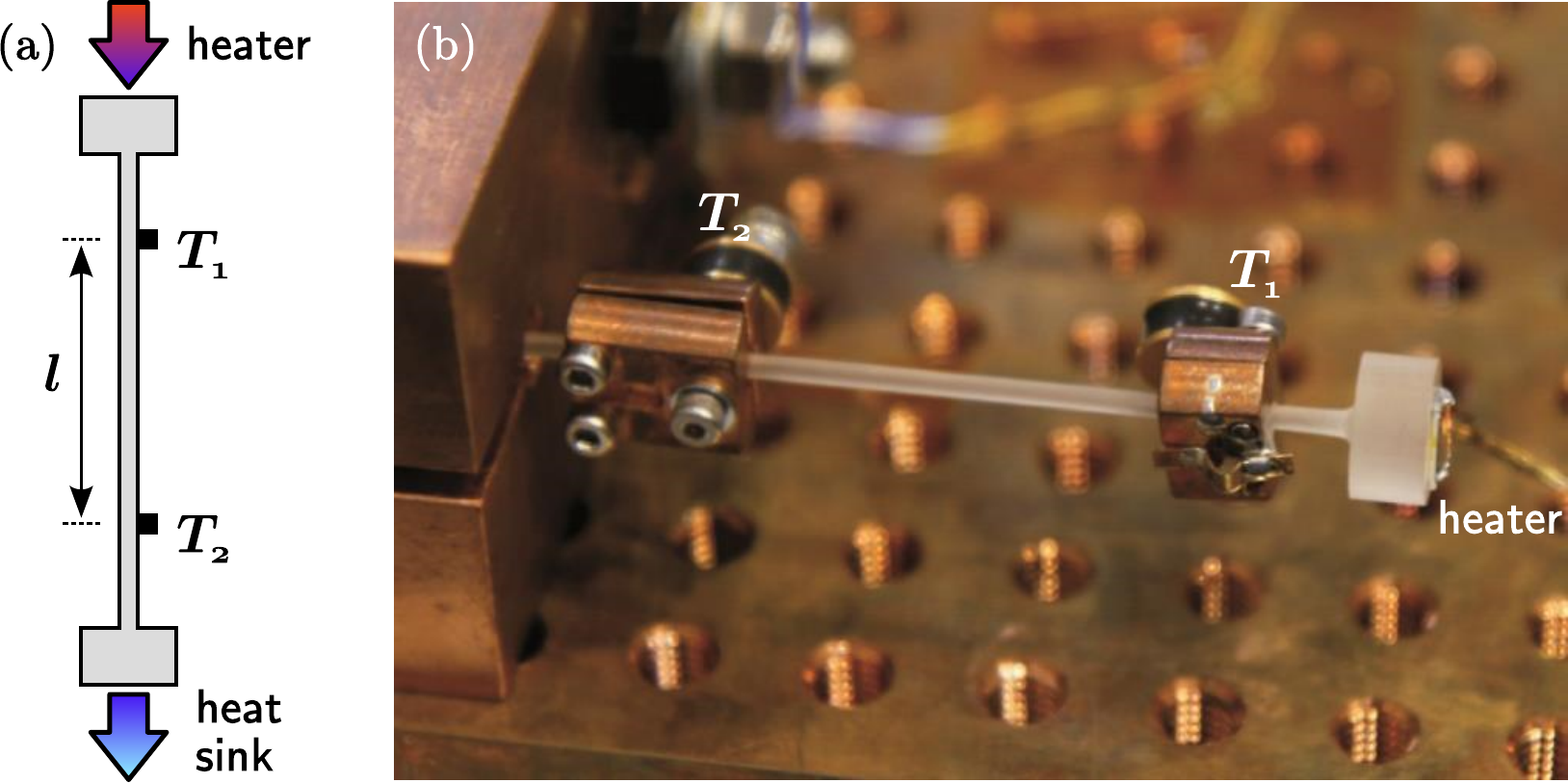}
		\caption{(a) Sketch of the setup employed for thermal conductivity measurements. A heater was fixed to one nail head, and two thermometers were used to measure the temperature gradient inside the fiber. The temperature of the heat sink could be controlled by using another heater, a thermometer was used for a feedback loop. (b) Photograph of the experimental setup.}
\label{fig:setup1}
\end{figure}
The experimental setup to measure the thermal conductivity is shown in Fig.~\ref{fig:setup1}. One end of the fiber was fixed to a copper heat sink that was later put in direct contact to a cryo-cooler unit (Sumitomo Heavy Industries, RP-062B), while an smd resistor with R\,=\,1\,k$\Omega$ was used as a heater, being firmly connected to the other nail head. Two calibrated temperature sensors (Lakeshore, DT-670) were brought into thermal contact with the fiber using customized clamps with a small contact surface to increase the precision of the measurement results (the length of the connection is 3\,mm, the distance between the sensors ranged between 47\,mm and 51\,mm in the different measurements). An aluminum heat shield around the full setup suppressed interactions via thermal radiation that could falsify the results of the measurement. All temperature sensors on the sample as well as the sample heater were connected to the outside via special low thermal conductivity cryo cable in order to minimize leaking heat through them. To adjust the temperature of the heat sink and thus of the fiber for a measurement of the thermal conductivity as a function of temperature, a Lakeshore temperature controller LS336 and a temperature sensor DT-670 were used.

For every measurement point it was first ensured that the system was in a thermal steady state. Then three different heating powers were applied to the heater fixed to the nail head, and the temperature gradient between the two sensors was measured, each time ensuring that the measurement was only done after a steady state was achieved. The heater power was chosen such that the maximal temperature difference $\Delta T_\textrm{max}$ between T$_1$ and T$_2$ was in the order of 200\,mK - 300\,mK which can be easily measured with the equipment in use. If heat radiation to the surroundings as well as heat conductivity through the electrical wiring can be neglected (so that, because of the thermal steady state, the power transmitted through the rod is equal to the heating power), the observed temperature difference is a linear function of the heating power:
\begin{equation}
\Delta P = \frac{A}{l}\kappa\Delta T,
\label{equation2}
\end{equation}
where $P$ is the electrical power, $A$ is the cross-section area of the fiber, $l$ is the separation length of the two temperature sensors and $\kappa$ is the thermal conductivity. The latter can hence be directly determined as
\begin{equation}
\kappa = \frac{l}{A}\frac{\Delta P}{\Delta T}.
\label{equation1}
\end{equation}
A 4-point readout, where two wires were used to feed the sensor with a constant current and two wires were used to read out the sensor voltage, was used measuring $P$ to increase the precision to better than 10\,$\upmu$W. The precisions for measuring the fiber diameter and sensor separation length were $\pm 0.05$\,mm and $\pm 0.5$\,mm, respectively. After one measurement set, the heat sink temperature was increased and the procedure repeated. The measurement temperature interval was 10\,K - 120\,K, the measurement procedure was controlled via a LabVIEW\footnote{Graphical programming environment, in our experiment used to realize a fully automated control of the measurement, http://www.ni.com/labview/} interface.

\begin{figure}
	\centering
\includegraphics[width=11cm]{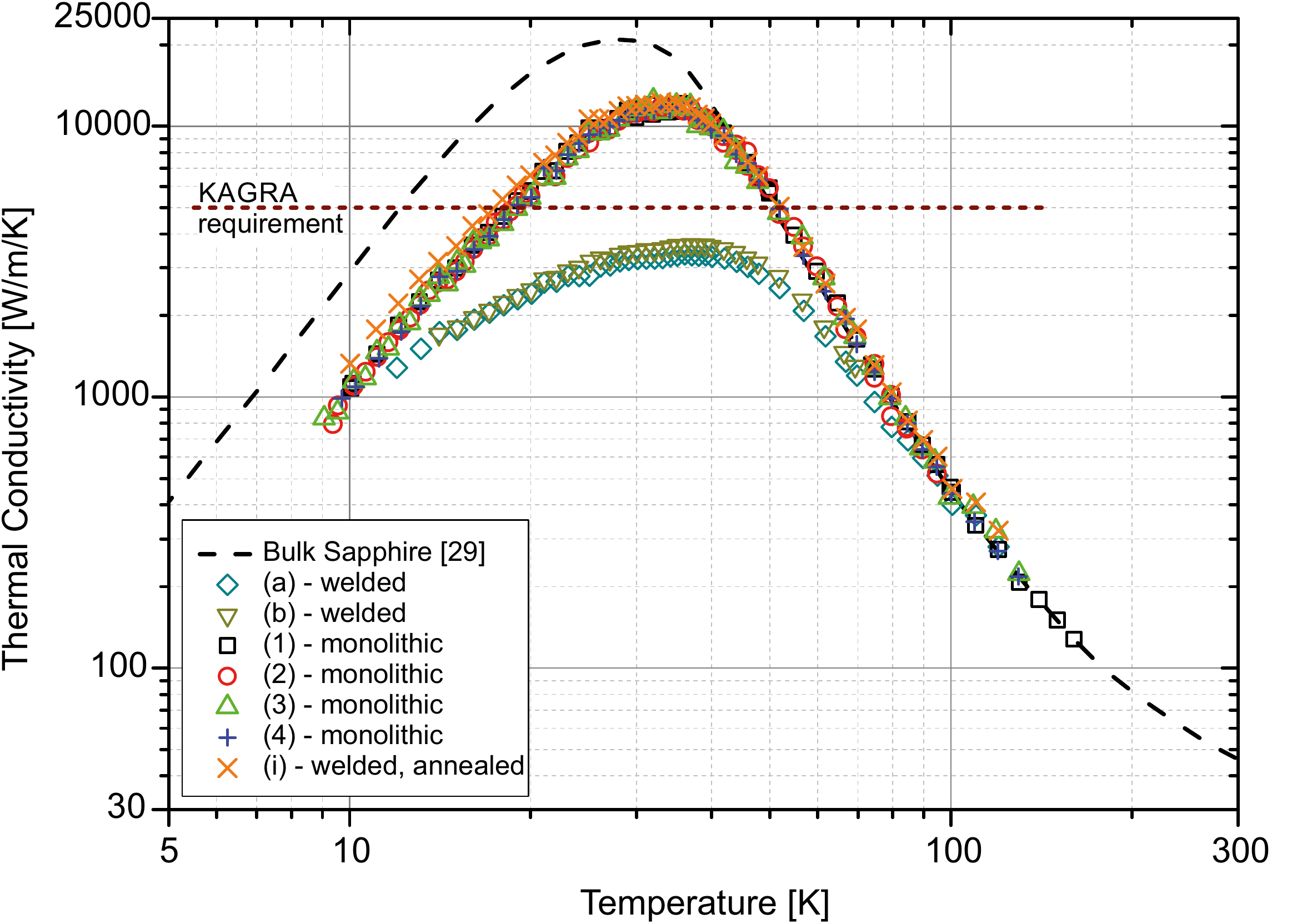}
		\caption{Thermal conductivities measured for different fiber types. (a) Fiber produced separately from three parts (monolithic nail head and short neck, a longer rod that was joined to the neck and another drilled nail head welded to the rod). Measurement over the longer part of the rod. (b) Same, but measurement over the joined part. (1-4) Monolithic fibers ground out of a single crystal. (i) Fiber produced separately, annealed at 1700\,°C for 60\,min and welded to two nail heads. Details are given in the text.}
\label{fig:Tcfibers}
\end{figure}

Figure~\ref{fig:Tcfibers} shows the thermal conductivities measured along the 1.6\,mm parts of the fibers. The names chosen for the traces correspond to the fiber names given above. The dashed black line shows bulk values derived from literature~\cite{touloukian}. In the temperature regime of KAGRA operation, the thermal conductivity of thin fibers will always lie below this curve because of the scattering of phonons at the fiber surfaces. The horizontal dotted line shows the KAGRA requirements of 5\,kW/m/K at 20\,K.\\
Traces (a\,-\,b) show the results obtained for the first fiber generation. Values of 2400\,W/m/K and 2480\,W/m/K were measured at 20\,K. The measurement error was estimated to a maximal value of $\pm 10\,\%$. The main contributions arise from measurement errors of the fiber diameter ($\pm 0.05$\,mm) and the distance between the two thermometers ($\pm 0.5$\,mm), while the heating power measurements contribute only to a very small amount (in the order of $0.5\%$) that can be neglected. The same holds true for the relative measurement of the temperature increase as a function of heater power. At fiber (a), $\kappa$ was measured over the long part of the rod only, while the measurement on fiber (b) included the transition between the monolithic neck and the joined rod. For this, one of the sensors was attached to the neck and the other sensor was placed on the rod, so that the temperature gradient over the connection area could be accessed. A comparison of traces (a) and (b) shows that the interface did not introduce a measurable heat resistance, thus proving the fabrication method to be in theory applicable. However, the thermal conductivity achieved at 20\,K with these samples lies a factor of two below the KAGRA requirements. The most plausible reason is the material used along with the bad surface quality and low diameter uniformity. According to the manufacturer, the Stepanov method was used to grow the fiber rod. Since material grown with this procedure is known to have bubbles and other defects close to the surface, usually the surface layers are removed after growth. In this case, however, rods with a diameter of 1.6\,mm were grown and no surface treatment was applied. Traces (1\,-\,4) show the results obtained for the fully monolithic fibers. Values of 5450\,W/m/K to 5850\,W/m/K were found at a temperature of 20\,K, satisfying the KAGRA requirements. The peak thermal conductivity was 11.8\,kW/m/K to 12.8\,kW/m/K. Finally, trace (i) presents the results for the welded fiber that underwent an annealing procedure. Although being with 6580\,W/m/K at 20\,K slightly higher, the effect of the annealing procedure on $\kappa$ is rather conservative.\\
Summarizing, the first set of measurements has shown that sapphire fibers of 1.6\,mm diameter satisfying and even exceeding the KAGRA requirements can be produced and that grinding the fibers separately and annealing them achieves the best results in terms of $\kappa$.

\section{Comparison between monolithic and welded fibers}
\label{sec:extraction}
\begin{figure}
	\centering
\includegraphics[width=10cm]{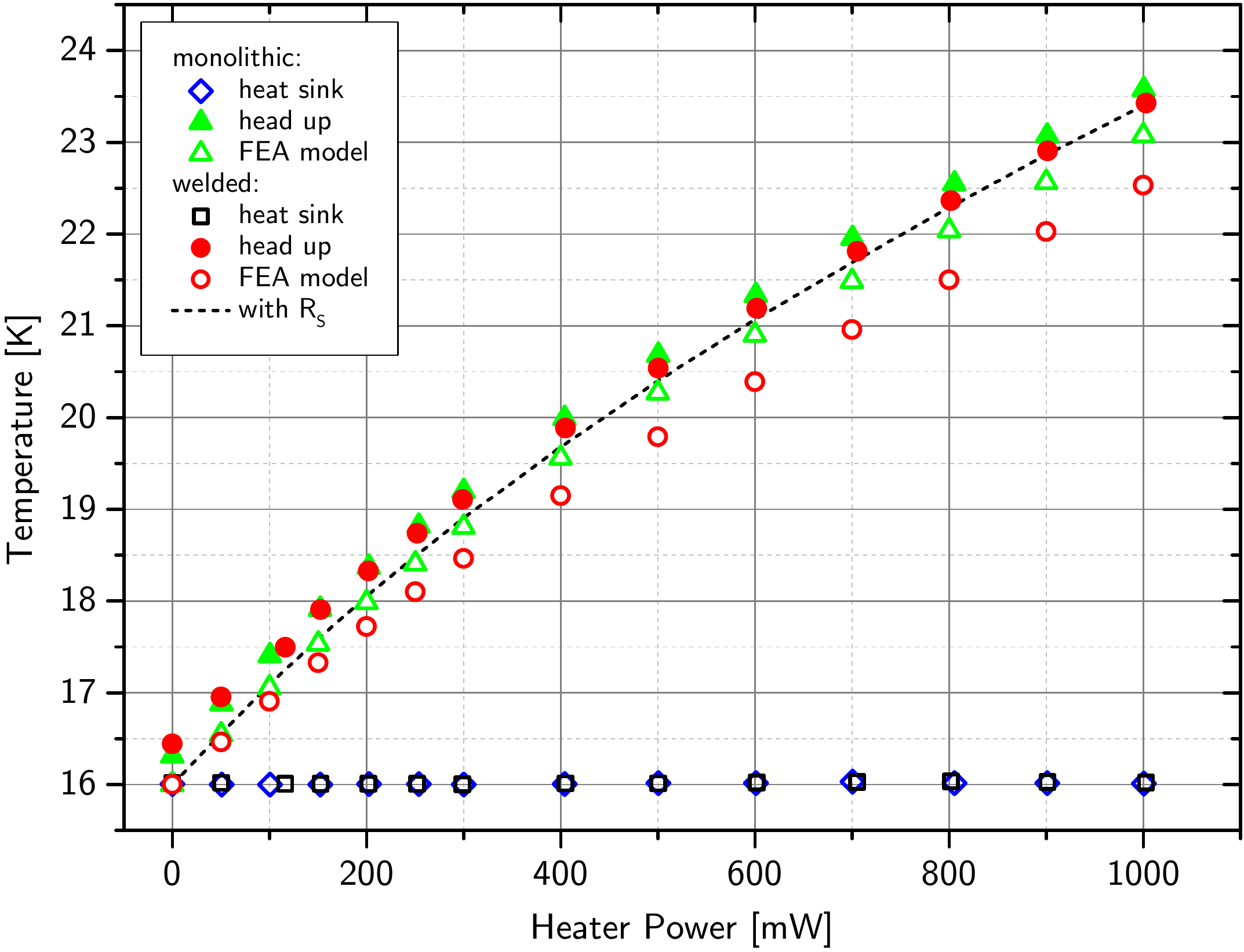}
		\caption{Temperatures of the nail head of a welded (red circles) and a monolithic (green triangles) fiber compared at different applied heater powers. The other fiber end was fixed to a heat sink that was kept at 16\,K (blue rhombs and black squares). The open red circles and green triangles show predictions from a finite element analysis (FEA) simulation. The thermal conductivity of the welded fiber was 15\,\% to 20\,\% higher than that of the monolithic one, so that from simulations a slightly lower head temperature was expected for identical heating powers. Dashed line: FEA model assuming a thin thermally resistive layer of $R_\mathrm{s} \approx 0.1\ldots0.2\,\mathrm{K cm^2 / W}$ at the location of the weld.}
\label{fig:heatextraction}
\end{figure}

As shown in Fig.\,\ref{fig:fiber}, the long thin fibers will be terminated by cylindrical or square shaped heads which then will be attached to the intermediate mass as well as to the test mass itself. Several techniques, including hydroxide catalysis bonding, a weld using a thin metal layer or direct optical contact are currently under investigation in view of their thermal conductivity, connection strength, the mechanical loss introduced to the system as well as the long-term behaviour. Although this still remains to be demonstrated, it seems, however, quite unlikely that any of the approaches chosen would introduce a limit concerning the extraction of heat from the test masses because of the large contact area available at the heads. However, while the heat transfer through the thin part of the fiber has already been shown to be within the KAGRA requirements, the weld between the fiber and the nail head might introduce an additional thermal resistance and thus narrow the cooling capability of the suspension. On the other hand, only the welding procedure makes it possible to use pre-polished nail heads and so to reach a surface flatness sufficient for hydroxide catalysis bonding or direct optical contact, making this manufacturing technique indispensable. For this reason, the welded connection between the thin fiber rod and the nail head was investigated in view of a possible thermal resistance in an independent measurement.\\
Two fibers, one of the monolithic type and a welded one, were compared. Again, an SMD-heater (R$\,=1000\,\Omega$) was placed on top of the nail head and along the fiber two temperature sensors were fixed, as already discussed in Fig.\,\ref{fig:setup1} for the previous measurements. Two additional temperature sensors were mounted on top and on the bottom of the nail head to monitor the temperature distribution and the thermal steady state of the nail head. 
The other end of the fiber was firmly connected to a copper heat sink. The heat sink was kept at 16\,K (likewise intended for the intermediate test mass in KAGRA). The thermal conductivity values of the fibers are shown in Fig.\,\ref{fig:Tcfibers}\,(1) and (i), please note that the values for fiber (i) were about 10\,\% - 20\,\% higher at lower temperatures.\\
To evaluate the heat extraction through the fiber, the heater power was gradually increased up to 1\,W and the temperatures of the nail heads were measured after a thermal steady state was reached. An immediate thermalization of the head was observed because of the high thermal conductivity. The results are shown in Fig.\,\ref{fig:heatextraction}. The blue rhombs and black squares show the temperature of the heat sink, being 16\,K. The solid green triangles and red circles show the temperature of the nail head as a function of heater power for the monolithic and the welded fiber, respectively. Finally, a simplified finite element analysis (FEA) model that merely takes into account the previously measured thermal conductivities of the two fibers was used for comparison with the observed values.
In the model the heater power was applied to the whole top nail head surface, which is a valid assumption considering the high thermal conductivity and the vanishing heat capacity of sapphire in the analyzed temperature range. Just as in the measurement setup, after a distance of 85\,mm the temperature of the fiber was fixed to 16\,K in the model, however, any thermal boundary resistance at the clamping point was neglected.
The results are shown in Fig.\,\ref{fig:heatextraction} as open green triangles and red circles. The deviation in expected heat extraction results from the different thermal conductivities of the two fibers, with the values for the welded fiber (i) being about 10\,\% - 20\,\% higher at lower temperatures.
For the monolithic fiber, a comparison of the measurement results and the FEA calculation shows only a slight deviation. This residual deviation originates from our very simplified and idealized model that assumes a perfect thermal coupling of all the parts of the fiber (which holds true for the monolithic fiber) and of the clamping to the heat sink.
By clamping the two fibers in the same way, the same thermal coupling was ensured in the experiment, allowing a relative statement about a possibly different behaviour of the two fibers.
In the case of the welded fiber the much stronger deviation from the model indicates that the welded connection additionally slightly reduces the heat extraction, so that the head heated up more than it would be expected. By adding a thin thermally resistive layer of $R_\mathrm{s} \approx 0.1\ldots0.2\,\mathrm{K cm^2 / W}$ in the FEA model at the location of the weld, a better agreement between model and experiment could be reached (dashed line in Fig.~\ref{fig:heatextraction}). Please note that the focus of the results reported here was a direct comparison between a monolithic and a welded fiber under identical experimental conditions in view of the possible heat extraction, while the FEA model was merely employed as a consistency check.

\section{Discussion and outlook}
\label{sec:conclusion}
The goal of the measurements reported in this paper was an analysis of prototype KAGRA sapphire suspension fibers in view of the possible heat extraction. Measurements on fibers ground out of a single crystal have yielded thermal conductivity values of up to 5.85\,kW/m/K at 20\,K. Even higher values of 6.6\,kW/m/K at 20\,K could be reached by annealing the fiber rod and thus improving the surface quality. These values exceed the value of $\kappa_\textrm{req}\geq\,$5000\,W/m/K that was drawn as a requirement for KAGRA. Thus our measurements demonstrated the extraction of heat deposited in the KAGRA test masses through absorbed laser light and heat radiation invading the cryostat through the view ports in general to be feasible with the current state of sapphire manufacturing and handling technology. Furthermore, a welded connection between the fiber rod and a head was analyzed and the results compared to a monolithic piece ground out of a single crystal. A simple FEA model, basing on the measured thermal conductivity values and assuming perfect thermal contact between fiber and heat sink as well as between thermometers and fiber, was developed and compared to the measurement results. While for the monolithic fiber the predictions from the numerical analysis matched the measurement results well within the accuracy of the measurement, for the welded fiber a minor difference was observed. A possible source for this deviation is that the welded connection introduces a heat resistance. Indeed, assuming a thin thermally resistive layer of $R_\mathrm{s} \approx 0.1\ldots0.2\,\mathrm{K cm^2 / W}$ in the FEA model at the location of the weld, a better agreement between model and experiment could be reached. It is, however, worth noting that the deviation might as well partially arise from a thermal boundary resistance at the clamping point of the fiber into the heat sink. There, a thin layer of indium was used to establish thermal connection, the contact area was about 8\,mm$^2$. Thus, our measurements define a lower boundary value of thermal load that can be extracted through sapphire fibers with welded heads. Up to 420\,mW could be extracted through the prototype fiber maintaining the head temperature below 20\,K and keeping the other end at 16\,K, mimicking the KAGRA case. Up-scaling this result to the full KAGRA fiber length of 300\,mm thus results in a maximal heat extraction of 120\,mW per fiber or of about 0.5\,W through a set of four fibers.\\
Please note that the numbers shown above merely constitute a lower boundary that, although already satisfying the requirements, can possibly be exceeded even further. In future, welded fibers with larger head dimensions as shown in Fig.~\ref{fig:fiber}\,(d) shall be produced and analyzed. This will allow us to determine the source of the deviation between FEA model and measurements discussed above, because limitations through a boundary heat resistance between the fiber head and the heat sink can be reduced by increasing the contact area. An area of up to 400\,mm$^2$ will be available. Furthermore, the area of the weld between fiber and head will be increased by a factor of two to three in order to access the influence of the weld area on the thermal conductivity.\\
At present, a concrete projection onto the final KAGRA case remains challenging because the total amount of heat deposited in the test masses will depend on parameters that in parts yet remain to be determined, as e.\,g.~the optical absorption in the test mass substrates and the dielectric coatings. Also, as shown in Fig.~\ref{fig:Tcfibers}, the thermal conductivities observed for individual fibers exhibit a variance by several hundreds W/m/K at 20\,K. Based on the results from fiber (i), the maximal heat extraction that can be expected in an ideal case not limited by thermal boundary resistances (red open circles in Fig.~\ref{fig:heatextraction}) would be 520\,mW for the prototype fiber, resulting in 150\,mW per fiber or 600\,mW per test mass when up-scaled to the KAGRA case. This value can be possibly slightly exceeded using pre-selected fibers with maximal thermal conductivity. From a broader view, other parameters that can be adapted are the fiber diameter and/or the designated operation temperature.\\
In summary, our measurements have shown that a first set of prototype sapphire fibers with a diameter of 1.6\,mm fulfills the KAGRA requirements in terms of thermal conductivity. Furthermore, an optimized design of the fiber heads was proposed to possibly further increase the heat flow through the weld area between head and fiber. 

\section*{Acknowledgements}
This work was supported in part by the Leading-edge Research Infrastructure Program of Japan, by the German Science Foundation DFG under contract SFB TR7 and by the EU under the ELiTES project (IRSES no 295153). AK was supported by the Japan Society for the Promotion of Science.

\end{document}